\documentclass{article}
\usepackage{spconf,amsmath,graphicx}

\usepackage{multicol, multirow}
\usepackage{adjustbox}
\usepackage{amssymb}
\usepackage{mathtools}
\usepackage{makecell, booktabs}
\usepackage{url, cite}
\usepackage{tabularx}
\newcolumntype{Y}{>{\centering\arraybackslash}X}
\usepackage{hyperref}
\usepackage{arydshln}
\usepackage{booktabs}
\hypersetup{
    colorlinks=true,
    linkcolor=purple,
    filecolor=magenta,      
    urlcolor=purple,
}
\usepackage{xcolor}
\newcommand{\newpara}[1]{\vspace{8pt}\noindent\textbf{#1}}

\title{Graph attentive feature aggregation \\for text-independent speaker verification}
\name{Hye-jin Shim$^{1\dagger}$, Jungwoo Heo$^{1\dagger}$\thanks{$^\dagger$Equally contributed.}, Jae-han Park$^{2}$, Ga-hui Lee$^{2}$, and Ha-Jin Yu$^{1*}$\thanks{$^*$Corresponding author.}}
\address{$^1$School of Computer Science, University of Seoul, $^2$KT Corporation}

\begin{document}
\ninept
\maketitle

\begin{abstract}
The objective of this paper is to combine multiple frame-level features into a single utterance-level representation considering pairwise relationship. 
For this purpose, we propose a novel graph attentive feature aggregation module by interpreting each frame-level feature as a node of a graph.
The inter-relationship between all possible pairs of features, typically exploited indirectly, can be directly modeled using a graph.
The module comprises a graph attention layer and a graph pooling layer followed by a readout operation. 
The graph attention layer first models the non-Euclidean data manifold between different nodes. 
Then, the graph pooling layer discards less informative nodes considering the significance of the nodes.
Finally, the readout operation combines the remaining nodes into a single representation. 
We employ two recent systems, SE-ResNet and RawNet2, with different input features and architectures and demonstrate that the proposed feature aggregation module consistently shows a relative improvement over $10\%$, compared to the baseline. 
\end{abstract}
\begin{keywords}
speaker verification, feature aggregation, attention, graph attention networks, deep learning
\end{keywords}

\section{Introduction}
\label{sec:intro}
Speaker verification (SV) can be used for various authentication scenarios where the identity of a given speech is compared to that of the claimed speaker. 
In SV, speaker representations are derived by first extracting the frame-level features and then aggregating them.
The network extracting the frame-level features is referred to as trunk network (e.g., convolutional neural network (CNN) and x-vector~\cite{jung2020improving,chung2020in, lin2020wav2spk, snyder2018x, tang2019deep, wu2020vector}).
After extracting the frame-level features, various techniques including gated recurrent network (GRU), learnable dictionary encoding (LDE) are used on top of the trunk network to aggregate frame-level features into a single utterance-level feature~\cite{li2017deep, jung2019short, zhang2019seq2seq, jung2019rawnet, jung2020improved,snyder2017deep, zhu2018self, okabe2018attentive, kye2021cross, cai2018exploring,xie2019utterance, jung2019spatial}.
The condensed utterance-level representation ultimately represents the entire character of an utterance, unlike speech recognition where each frame-level representation is of worth.
Hence, feature aggregation plays an essential role and a number of studies in the recent literature have focused on developing feature aggregation methods.

A group of studies employed recurrent layers for feature aggregation. 
These studies modeled the frame-level representations in sequential order~\cite{li2017deep, jung2019short, zhang2019seq2seq, jung2019rawnet, jung2020improved}.
Another group of research integrated frame-level features using LDE \cite{cai2018exploring, xie2019utterance, jung2019spatial}.
Similar to the recurrent layers, the LDE layer aggregates features over time whereas it further utilizes statistics.
Sequential modeling, a foundation of the above two approaches, has been used in SV based on its success in language processing and speech recognition, where the sequential order could be a prior knowledge.
However, recent studies found that sequential information may not be the key, especially in text-independent SV (TI-SV)~\cite{okabe2018attentive, desplanques2020ecapa}.

Other groups of researches exploited attention-based approaches-~\cite{snyder2017deep, zhu2018self, okabe2018attentive, kye2021cross}, which exclusively emphasize important features regardless of their sequence and have recently become the mainstream in TI-SV.
These approaches assign an attention weight in the form of a scalar to each frame-level feature and then perform weighted summation (also with statistics as in \cite{okabe2018attentive}). 
Each attention weight for a feature is derived via a dot product between a feature and a projection vector. 
A softmax non-linearity is applied to attention weights to exclusively emphasize which frame is more important, however, they are compared indirectly without modeling their relationships. 

The field of graph neural networks is recently getting attention by incorporating the advantages of graph structures and deep neural networks~\cite{goyal2018graph, wu2020comprehensive}.
Interpreting high-dimensional representations as nodes of a graph, graph neural networks can model the non-Euclidean data manifold within nodes, including {\em inter-node relationship}.
Particularly, recent architectures such as graph convolutional network~\cite{kipf2016semi} and graph attention network (GAT)~\cite{velivckovic2017graph} combined with various graph pooling layers~\cite{gao2019graph, lee2019self, yuan2020structpool} have demonstrated competitive performance even in image and audio domains~\cite{jung2021graph, tak2021graph, tak2021end}. 

Inspired by the recent success of graph neural networks, we argue that the feature aggregation could be further improved by explicitly comparing and modeling the inter-relationship between frame-level features.
The feature aggregation is a task to express characteristics relative to other features and condense them into a single representation.
Hence, the inter-relationship between features extends the ability of its expressiveness and could be the vital information itself.
To this end, we propose a novel feature aggregation module to model inter-relationship by interpreting each frame-level feature as a node of a graph.
The proposed module leverages the correlation of all possible pairs of nodes to obtain an utterance-level representation.
In other words, entire frame-level features are aggregated considering their inter-relationships in addition to their intrinsic characteristics by utilizing a graph.
Specifically, the aggregation involves a GAT, a graph pooling layer, and a readout operation~\cite{velivckovic2017graph,gao2019graph}.
The GAT assigns different weights to each node pairs (edges) and models their relationships.
Then, the graph pooling layer discards less informative nodes. 
Finally, a single utterance-level representation is derived using the readout operation. 

\begin{figure*}[ht!]
\begin{center}
    \centering
    \includegraphics[width=0.65\linewidth]{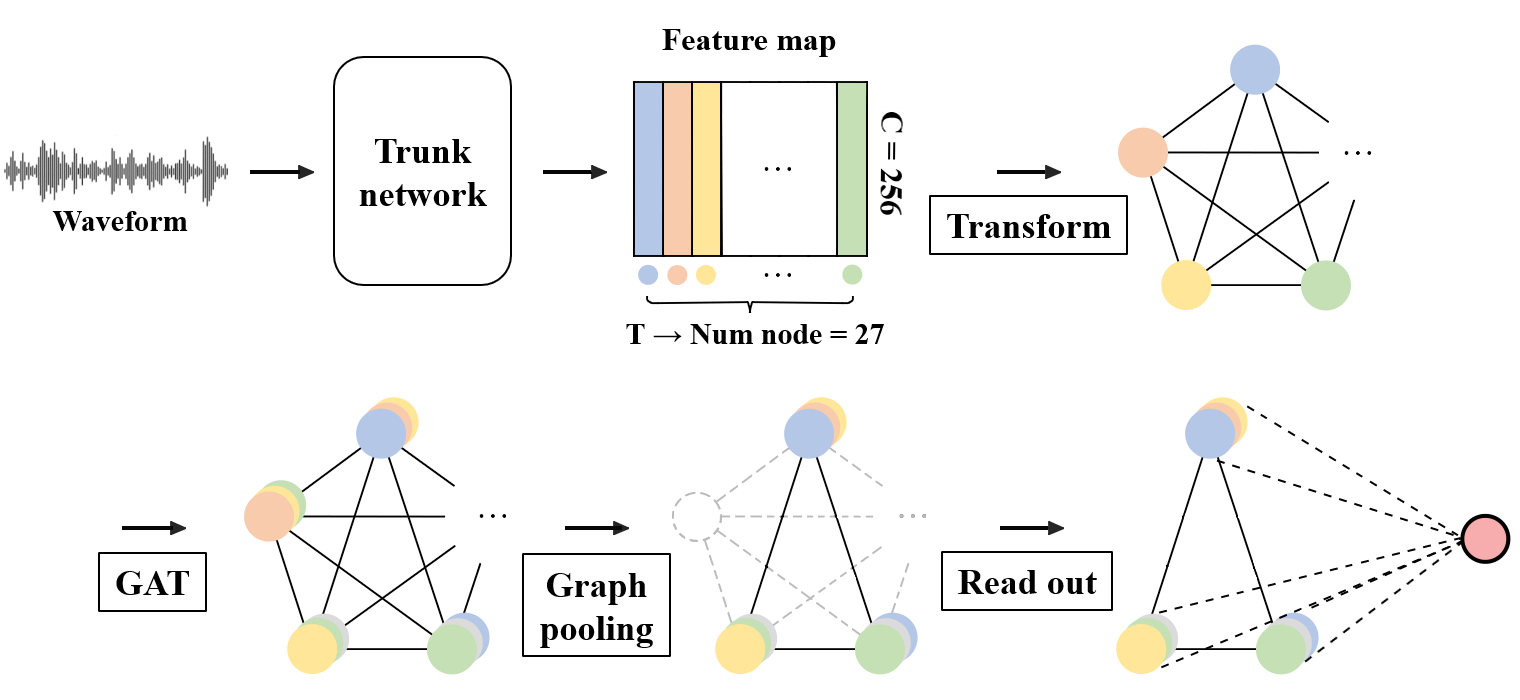}
\vspace{-1.5em}
\caption{
  Proposed graph attentive feature aggregation module. After extracting frame-level features using a trunk network, a graph is constructed using each sequence element as a node. 
  Each node can have a dimensionality equal to either {\em the number of filters} (raw waveform, 1D-CNN) or {\em the number of filters $\times$ the number of frequency bins} (mel-filterbank, 2D-CNN).
  Graph attention network first aggregates nodes features, and then the graph pooling reduces the number of nodes. 
  The readout function integrates nodes into a single representation (best viewed in color).}
\label{fig:overall}
\end{center}
\vspace{-1.5em}
\end{figure*}

The effectiveness of the proposed feature aggregation framework is validated using two strong baselines, SE-ResNet~\cite{heo2020clova} and RawNet2 \cite{jung2020improved}, comprising different input features and architectures.
Experimental results on the VoxCeleb datasets~\cite{nagrani2017voxceleb,chung2018voxceleb2} demonstrate more than 10\% consistent improvement over the corresponding baseline.
Using the proposed graph aggregation module, we could compose more lightweight, yet better performing models compared to self-attentive pooling and GRU-based aggregations.

\section{graph attentive feature aggregation}
\label{sec:proposed}
In this section, we introduce the graph attentive feature aggregation module.
The proposed module is located after the extraction of frame-level features using a trunk network (RawNet2 or SE-ResNet).
The proposed approach consists of three components: i) graph attention layer, ii) graph pooling layer, and iii) readout.
We first describe how we formulate a graph from a feature map and perform GAT to integrate the overall features. 
Then, we explain how the graph pooling layer discards less informative nodes and the readout combines the results into a single representative feature.
The overall scheme is illustrated in Figure~\ref{fig:overall} using RawNet2 (the scheme is identical for SE-ResNet except the dimensionality of each node).

\subsection{Graph attention layer}
We first formulate a graph using frame-level features extracted from a trunk network.
Specifically, we interpret each sequence element as a node of a graph; for RawNet2, each node has a dimensionality equal to {\em the number of filters} and for SE-ResNet, the dimensionality equals to {\em the number of filters $\times$ the number of frequency bins}.
Note that we design a fully-connected (complete) graph where all possible edges exist between pairs of nodes. 
Nodes can be directly compared considering all combinations/relationships because all nodes have bi-directional edges. 
Hence, entire frame-level features are aggregated considering their inter-relationships as well as intrinsic characteristics of their own by utilizing the graph attention layer.

Let $\mathbb{G}$ be a complete graph comprising $N$ nodes with the $F$ dimensional features of each node.
A set of nodes in $\mathbb{G}$ is defined as $\mathbf{x}$ $\in$ $\mathbb{R}^{N \times F}$ and each nodes are represented as row vectors $x_1$, $x_2$, $\cdots$, $x_N$.
In the GAT layer, $\mathbf{x}$ is first projected into $F'$ dimensional space using a matrix multiplication with $W \in \mathbb{R}^{F \times F'}$ resulting in $\mathbf{n'} \in \mathbb{R}^{N \times F'}$ (Equation (1)).
Then, we calculate attention scores where an attention score $a_{ij}$ between $i$-th node $n'_i$ and $j$-th node $n'_j$ ($n'_i$, $n'_j$ $\in \mathbb{R}^{F'}$, $i \neq j$) is calculated through Equation (2) and (3). 
We derive $e_{ij}$, by first concatenating two nodes $n'_i$ and $n'_j$ and then projecting to a scalar value through a dot product with $\gamma \in \mathbb{R}^{2F' \times 1}$, a learnable parameter, followed by a Leaky ReLU non-linearity~\cite{maas2013rectifier}.
Using the calculated attention scores, the GAT performs self-attention on node $n'_i$ as in Equation (4).
The GAT is applied to every node $\mathbf{x}$ transforming $\mathbf{n'}$ to $\mathbf{n}$ and then the output $\mathbf{n}$ is fed to a graph pooling layer.

The GAT process can be described as follows:
\begin{gather}
    \mathbf{n'} = \mathbf{x} W,\\
    e_{ij} = LeakyReLU(\gamma \cdot \mathsf{concat} ({n'}_i, {n'}_j)), \\
    {a}_{ij} = \mathsf{softmax}(e_{ij}) = \frac{\mathsf{exp}(e_{ij})}{\sum_{t=1}^N \mathsf{exp}(e_{it})},\\
    {n}_i = \sum_{j=1}^N {a}_{i j} {n'}_j.
\end{gather}

\begin{table}[]
    \caption{
        SE-ResNet architecture applying the proposed graph aggregation method.
        ReLU and batch normalization layers are omitted.
        Squeeze and excitation~\cite{hu2018squeeze} is performed after all convolutional blocks.
        The numbers denoted in brackets refer to \textbf{[size$\times$ size, filters]} and those in the output refer to \textbf{(filters, frequency, time)}.
        The output of the last fully-connected layer is used as the speaker embedding.
    }
    \begin{tabular}{c c c}
    \midrule
        \textbf{layer name} &   \textbf{Filters}     & \textbf{Output} \\ 
    \midrule
        conv1       & 3$\times$3, 32, stride 1   & (32$\times$40$\times$T)\\
    \midrule
        conv2 & 
        $\left [
            \begin{tabular}{c}
            3$\times$3, 32\\
            3$\times$3, 32\\
            \end{tabular}
        \right ]\times3$, stride 1
        & (32$\times$40$\times$T) \\
    \midrule
        conv3 & 
        $\left [
            \begin{tabular}{c}
            3$\times$3, 64\\
            3$\times$3, 64\\
            \end{tabular}
        \right ]\times4$, stride 2
        & (64$\times$20$\times$T/2) \\
    \midrule
        conv4 & 
        $\left [
            \begin{tabular}{c}
            3$\times$3, 128\\
            3$\times$3, 128\\
            \end{tabular}
        \right ]\times6$, stride 2
        & (128$\times$10$\times$T/4) \\
    \midrule
        conv5 & 
        $\left [
            \begin{tabular}{c}
            3$\times$3, 128\\
            3$\times$3, 128\\
            \end{tabular}
        \right ]\times3$, stride 2
        & (128 $\times$ 5 $\times$T/8) \\
    \midrule
    aggregation &   GAT         & (640)\\
    \midrule
    fc          &   FCN, 256    & (256)\\
    \midrule
    \end{tabular}
    \label{tab:seresnet}
    \vspace{-1.5em}
\end{table}

\begin{table}[t]
 \caption{Application of proposed graph module to SE-ResNet and RawNet2. SAP in SE-ResNet and GRU in RawNet2 are both replaced by GAT. } %
  \centering
  \label{tab:gat}
  \begin{tabular}{l c c c}
  \toprule
  \textbf{Feature extractor} & \textbf{Aggregation} & \textbf{\# Params} &\textbf{EER (\%)} \\
  \toprule
  SE-ResNet & SAP & 6.0M & 1.98 \\
  SE-ResNet & GAT & 5.4M &\textbf{1.86} \\
  \midrule
  RawNet2 & GRU & 13.2M & 2.48 \\
  RawNet2 & GAT & 9.9M & \textbf{2.23} \\
  \bottomrule
  \end{tabular}
  \vspace{-1.5em}
\end{table}

\begin{table}[t]
 \caption{
   Comparison of applying different pooling ratios in graph pooling layer. A higher pooling ratio means more nodes remain.}
  \centering
  \label{tab:gpool}
  \begin{tabular}{l c c}
  \toprule
  \multirow{2}{*}{\textbf{Pooling method}} & \multicolumn{2}{c}{\textbf{EER (\%)}} \\
  & SE-ResNet & RawNet2 \\
  \toprule
   w/o gPool & 1.86 & 2.23 \\
   Top 11\% gPool & 2.00 & 2.39 \\
   Top 33\% gPool & 1.91 & 2.17 \\
   Top 80\% gPool & \textbf{1.75} & \textbf{2.15} \\
  \bottomrule
  \end{tabular}
  \vspace{-1.5em}
\end{table}

\begin{table}[t]
 \caption{Comparison of various readout methods. ``Combine by concat’’ indicates that the results of summation, standard deviation (std), minimum, and maximum are concatenated by reducing the output dimensions in one-quarter.}
  \centering
  \label{tab:nodeagg}
  \begin{tabular}{l c}
  \toprule
  \textbf{Operation} & \textbf{EER (\%)} \\
  \toprule
   Mean & 2.38 \\
   Sum & \textbf{2.23} \\
   Max & 2.3 \\
   Combine by concat  & 2.38 \\
  \bottomrule
  \end{tabular}
  \vspace{-1.5em}
\end{table}

\subsection{Graph pooling layer and readout}
We utilize a graph pooling layer followed by the readout operation to obtain an utterance-level feature. 
Similar to pooling layers in convolutional neural networks, graph pooling can reduce the size of features to enable high-level feature encoding and receptive field enlargement.
Specifically, the graph pooling layer reduces the original graph into a sub-graph by removing less informative nodes as depicted in Figure~\ref{fig:overall}. 
For the graph pooling, we exploit gPool introduced in the Graph U-Net architecture~\cite{gao2019graph}.
There are various other graph pooling methods, but many of them assume the hierarchical architecture of the graph.
As it is difficult to define a hierarchical relationship between frame-level features in audio data, we adopt gPool, which does not require the graph to have a hierarchical structure.
gPool consists of the following processes: projection, top-$K$ node selection, and a gating mechanism.

A set of nodes in the graph pooling layer $l$ is represented as ${\mathbf{n}}^{l}$.
Note that the adjacency matrix is omitted because we utilize a complete graph which is different from the original paper of~\cite{gao2019graph}.
In the projection stage, gPool obtains score vector $\mathrm {y}$ by estimating the scalar projection values of each node, using a trainable projection vector  $\mathrm{p} \in \mathbb{R}^{F'}$.
In the top-$K$ node selection, the operation of node ranking ($\mathsf{rank}$) outputs indices of the $K$-largest values ($\mathrm{index}$) based on calculated score vector $\mathrm{y}$.
The score vector of selected nodes, $\mathrm{\tilde {y}}$, can be obtained by applying $\mathsf{sigmoid}$ to each element in the
extracted scalar projection vector. 
In the gate stage, element-wise product $\odot$ is conducted between the pooled nodes $\mathbf{\tilde{n}}^l$ and the selected node scores $\mathrm{\tilde {y}}$.

Formally, gPool can be denoted as follows:
\begin{equation} 
\begin{aligned}
    &\mathrm{y}=\mathbf{n}^{l} \mathrm{p}^{l} / \lVert{\mathrm{p}}^{l}\lVert, \\
    &\mathrm{index}=\mathsf{rank}(\mathrm{y}, K), \\
    &\mathrm{\tilde y}=\mathsf{sigmoid}(\mathrm{y}(\mathrm{index})), \\
    &\tilde {\mathbf{n}}^{l} = \mathbf{n}^{l}(\mathrm{index}, :) \\
    &\mathbf{n}^{l+1} = \tilde {\mathbf{n}} ^{l} \odot \mathrm{\tilde y} \\
\end{aligned}
\end{equation}

After the graph pooling, we conduct the readout operation, which combines the processed nodes into a single node.
We explore various types of readout operation where the best result is obtained using the summation of nodes:
\begin{equation}
\begin{aligned}
    &U=\sum_{k=1}^K n_k^{l+1}.\\ 
\end{aligned}
\end{equation}

Here, $U$ is the aggregated utterance-level feature and $K$  nodes $n_k$ are integrated to $\mathbf{n}$. 
The results of various readout mechanisms can be found in Section~\ref{sec:result}.

\begin{table*}[t]
 \caption{Results of comparison with recent state-of-the-art systems. The two baselines of this study are each compared with various architectures that use the same input feature.}
  \centering
  \label{tab:sota}
  \begin{tabular}{l c c c c c c c }
  \toprule
    & Input Feature & Front-end & Aggregation & Loss & EER (\%) \\
  \toprule 
  Chung \textit{et al.} \cite{chung2020in} & Spec-257 & Thin ResNet-34 & SAP & AP & 2.21 \\
  Yu \textit{et al.} \cite{yu2019ensemble} & Spec-512 & ResNet-50 & TAP & EAM-Softmax & 2.94 \\
  Liu \textit{et al.}
 \cite{liu2021effective} & MFB-40 & Dense-Residual & ABP & Softmax & 2.54 \\
  Jung \textit{et al.} \cite{jung2020improving} & MFB-40 & Fast ResNet-34 & LDE & NP + Softmax & 1.98 \\
  Kye \textit{et al.} \cite{kye2021cross} & MFB-40 & Fast ResNet-34 & CAP & NP + Softmax & 1.88 \\
   Ours & MFB-40 & SE-ResNet & SAP & AP + AAM-Softmax & 1.98\\
  \textbf{Ours(Proposed)} & MFB-40 & SE-ResNet & GAT & AP + AAM-Softmax & \textbf{1.75}\\
  \bottomrule
  Lin \textit{et al.} \cite{lin2020wav2spk} & \multirow{4}{*}{Raw waveform} & wav2spk & Gating + SP & AM-Softmax & 3.00 \\
  Zhu \textit{et al.} \cite{zhu2021vector} &  & Y-vector & SP & AM-Softmax &  2.60\\
  Jung \textit{et al.} \cite{jung2020improved} &  & RawNet2 & GRU & Softmax & 2.48\\
  \textbf{Ours(Proposed)} &  & RawNet2 & GAT & AM-Softmax & \textbf{2.15}\\
  \bottomrule
  \end{tabular}
  \vspace{-1.5em}
\end{table*}

\begin{table}[t]
 \caption{Results of additional experiments. The first row shows the result of applying GAT and GRU sequentially. The lower two rows show the result of applying two GAT layers with different pooling architectures.}
  \centering
  \label{tab:featext}
  \begin{tabular}{l c}
  \toprule
  \textbf{System} & \textbf{EER (\%)} \\
  \toprule
  GAT - GRU & 4.16 \\ 
  \midrule
  2GAT - Global architecture & 2.66 \\
  2GAT - Hierarchical architecture & 2.55 \\
  \bottomrule
  \end{tabular}
  \vspace{-1.5em}
\end{table}

\section{Experiments}
\subsection{Datasets}
All experiments are performed using the VoxCeleb1\&2 datasets~\cite{nagrani2017voxceleb, chung2018voxceleb2}. 
We train the model using the development subset of VoxCeleb2 that includes the utterances from 5,994 speakers.
Then, the evaluation is performed using the original trial that uses VoxCeleb1's evaluation subset.

\subsection{Implementation details}
Both baselines using a mel-filterbank and a raw waveform are implemented based on the PyTorch framework \cite{paszke2019pytorch}.
Graphs are implemented using Deep Graph Library \cite{wang2019deep}.
We adjust the number of attention heads and use 16 heads in RawNet2 and 32 heads in SE-ResNet based on empirical results.

\newpara{SE-ResNet.} 
Our mel-filterbank baseline is most similar to the architecture of~\cite{heo2020clova}, but overall details are adjusted.
We use 40-dimensional mel-filterbank features extracted with 1,024 point FFT and a hamming window of width 25ms and step 10ms.
The mel-filterbank baseline is optimized using Adam optimizer which uses a learning rate of 0.001, and the learning rate is decreased by 5\% in each epoch. 
 We use two types of loss functions, additive angular margin softmax (AAM-softmax)~\cite{deng2019arcface} and angular prototypical loss (AP)~\cite{wang2017deep} to consider inter-class as well as intra-class covariance. 
We employ a margin of 0.3 and a scale of 30 for AAM-softmax. 
For AP, we use a mini-batch size of 200, where each mini-batch contains two utterances per speaker.
The system is trained for 100 epochs. 
Our overall architecture applying the proposed method is presented in Table~\ref{tab:seresnet}.

\newpara{RawNet2.}
RawNet2 \cite{jung2020improved} is an end-to-end system that is fed by raw waveforms directly without preprocessing techniques.
For mini-batch construction, utterances are either cropped or duplicated (concatenated) into 59,049 samples ($\approx$ 3.69s) in the training phase, following~\cite{jung2020improved}.
In the evaluation phase, no adjustments are made to the length.
We modify several details from RawNet2~\cite{jung2020improved} in the process of adjusting the GAT as follows: 
i) exclude sinc-conv layer, as it slows down training time in spite of showing similar performance,
ii) replace softmax with  additive margin softmax (AM-softmax)~\cite{wang2018cosface},
iii) reduce the dimensions by half for the last fully-connected layer before speaker embedding (1024 dimensions to 512 dimensions).

\section{Results}
\label{sec:result}
\newpara{Application of GAT.} 
In Table~\ref{tab:gat}, we first address the effectiveness of the graph attention layer on our two baselines: SE-ResNet and RawNet2. 
Here, the graph pooling is temporarily excluded and summation readout operation is applied.
Adoption of the graph attention layer reduced the EER relatively by 6\% and 10\%, respectively.
In addition, we confirm that the number of parameters also decreases (by almost 25\% in the case of RawNet2 (13.2M to 9.9M)).

\newpara{Graph pooling.} 
In Table~\ref{tab:gpool}, we explore the effect of applying the graph pooling layer between the graph attention layer and the readout operation with diverse pooling ratios.
The denoted percentage of pooling refers to the remaining ratios of nodes; higher ratios mean that fewer nodes have been removed. 
Top 80\% pooling removing 20\% of irrelevant nodes shows the best results where the relative improvement compared to the baseline becomes 11.6\% and 13.3\%.
Through a comparison of the three different ratios of pooling, we find that discarding too many nodes worsens the performance (11\% pool) but removing a few of the most irrelevant nodes improves the performance (80\% pool).  
We thus conclude that the discriminative information is well combined into the majority of nodes (80\% in our case) via the GAT layer. 

\newpara{Comparison of readout operations.} 
In Table~\ref{tab:nodeagg}, we further compare several readout operations using the RawNet2 baseline.
We find that the proposed graph-based aggregation shows consistent effectiveness, regardless of the adopted readout approach because all node aggregation methods performed better than GRU.
Among various methods, summation achieved the best result.

\newpara{Comparison with state-of-the-art systems.} 
In Table~\ref{tab:sota}, we compare the proposed system with other recent state-of-the-art systems in the literature that adopt various feature aggregation methods.
None of the systems apply data augmentation techniques. 
In both input features, the proposed models using the proposed graph aggregation module demonstrated the best performance, showing the effectiveness of the proposed approach.
Given that we did not use data augmentation methods, there is room to improve the system using various data augmentation.

\newpara{Additional experiments.} 
Table~\ref{tab:featext} addresses two additional experiments using the RawNet2 baseline. 
First, we explore whether adopting a GRU after the GAT is beneficial and show the result in the first row. 
To maintain the overall complexity of the model, we omitted the last residual block and then placed the GAT and the GRU in sequence. 
However, this worsens the performance, as presented in the first row.
In our analysis, this shows that applying GRU directly to the GAT's output is not effective, because the GAT's output is not sequential. 

Second, we explore using two GAT layers with different pooling architectures proposed in ~\cite{lee2019self} and denote the result in second and third rows.
Both architectures adopt two GAT layers in sequence. 
The global architecture concatenates two GAT layers' output and feeds it to the graph pooling layer followed by the readout, whereas hierarchical readout performs graph pooling and readout after each GAT layer and then adds them element-wisely.
Through experiments, we found that both modified pooling architectures did not bring further improvements.

\section{Conclusions}
In this paper, we proposed a graph feature aggregation method for TI-SV.  
Utilizing a GAT, graph pooling layer, and readout operation, we directly modeled the inter-relationship between entire frame-level features, which is partially or indirectly utilized in the existing methods.
As this is the first work employing graph neural networks for feature aggregation, we also explored various configurations to optimize the system.
Consistent improvements over the baselines with different aggregation modules demonstrate the effectiveness of the proposed approach. 
\vspace{0.5em}
\newline
\textbf{Acknowledgement.} 
We would like to thank Jee-weon Jung at Naver Coporation for his help with the conceptualization and editing.

\bibliographystyle{IEEEbib}
\bibliography{shortstrings,reference}
\end{document}